\documentclass[pra,superscriptaddress,showpacs,preprint]{revtex4}
\usepackage{epsfig}
\usepackage{graphicx}
\usepackage[latin1]{inputenc}

\newcommand{\EVRY}{D\'epartement de Physique et Mod\'elisation,
Universit\'e d'Evry Val d'Essonne\\
Boulevard F. Mitterrand, 91025 Evry, France}
\newcommand{\LKB}{Laboratoire Kastler Brossel, Universit\'e Pierre et Marie Curie\\
T12, Case 74, 4 place Jussieu, 75252 Paris, France}

\begin{document}
\title{Vibrational spectroscopy of $\mbox{H}_2^+$: \\
hyperfine structure of two-photon transitions }

\author{Jean-Philippe Karr}
\email{karr@spectro.jussieu.fr}
\affiliation{\LKB}
\affiliation{\EVRY}

\author{Franck Bielsa}
\affiliation{\LKB}
\affiliation{\EVRY}

\author{Albane Douillet}
\affiliation{\LKB}
\affiliation{\EVRY}

\author{Jofre Pedregosa Gutierrez}
\affiliation{\LKB}
\affiliation{\EVRY}

\author{Vladimir I.~Korobov}
\affiliation{Joint Institute for Nuclear Research, 141980, Dubna, Russia}

\author{Laurent Hilico}
\affiliation{\LKB}
\affiliation{\EVRY}

\date{\today}
\begin{abstract}
We present the computation of two-photon transition spectra between ro-vibrational states of the H$_2^+$ molecular ion, including the effects of hyperfine structure and excitation polarization. The reduced two-photon matrix elements are obtained by means of a variational method. We discuss the implications of our results for high-resolution spectroscopy of H$_2^+$.
\end{abstract}
\maketitle
%%%%%%%%%%%%%%%%%%%%%%%%%%%%%%%%%%%%%%%%%%%%%%%%%%%%%%%%%%%%%%%%%%%%%%%%%%%%%%%%%%
\section{Introduction}
The H$_2^+$ ion is the simplest stable molecule. It plays an important role, both as a benchmark system for detailed studies of molecular energy levels~\cite{leach1995}, and in astrophysics. However, there have been very few investigations concerning high-resolution spectroscopy of H$_2^+$.   Radiofrequency spectroscopy of the hyperfine structure has been performed on H$_2^+$ ions trapped in a Paul trap~\cite{jefferts1969}. Rotational and ro-vibrational transitions close to the dissociation limit were investigated using microwave and laser spectroscopy on an ion beam~\cite{carrington1996,critchley2001}. The scarcity of experimental studies is mainly due to the fact that H$_2^+$, being homonuclear, does not have a dipole-allowed rotational or vibrational spectrum (except in a small region close to the dissociation limit, where the $1s \sigma_g$ and $2p \sigma_u$ electronic curves overlap).

Two-photon ro-vibrational transitions are nevertheless allowed, and Doppler-free two-photon spectroscopy was proposed a few years ago as a promising new method for determination of the electron-to-proton mass ratio $m_e/m_p$~\cite{hilico2000,koelemeij2007}. Transition probabilities between $L=0$ states were computed in~\cite{hilico2001} (where $L$ is the total orbital angular momentum quantum number), demonstrating the feasibility of two-photon spectroscopy using $\Delta v = 1$ transitions around 8-12 $\mu$m. Among these, transitions lying in the spectral range of CO$_2$ lasers (9-10 $\mu$m) are especially attractive for frequency metrology, because of their high output power and stability. Even if there is no coincidence of H$_2^+$ transitions with the CO$_2$ lines, a CO$_2$ laser can be used as a frequency reference for a tunable quantum cascade laser (QCL)~\cite{bielsa2007}. We have built an experiment designed to probe the $(v=0,L=2) \rightarrow (v'=1,L'=2)$ transition at 9.166 $\mu$m~\cite{karr2007}.

The aim of this paper is to present the computation of two-photon transition spectra in H$_2^+$, including the effects of hyperfine structure. In Sec.~\ref{firstpart}, the hyperfine effective Hamiltonian obtained in~\cite{korobov2006} is diagonalized, and the hyperfine states are written explicitly. In Sec.~\ref{secondpart}, we recall the theory of two-photon transition probabilities. The transition matrix elements between hyperfine states are expressed as a function of reduced matrix elements involving only the orbital part of the wave functions, which are calculated using the same variational method  as in~\cite{korobov2006}. In order to avoid huge data, only the spectra of the transitions $(v\!=\!0,L) \rightarrow (v'\!=\!1,L)$ with $0 \leq L \leq 3$ are presented \cite{suppl}. One reason for this choice is that the H$_2^+$ hyperfine structure is essentially determined by the value of $L$. Moreover, the considered $L$ values are the only one which are significantly populated when H$_2^+$ ions are created by electron impact ionization on H$_2$ at room temperature, and the frequencies of these transitions are sufficiently close to a CO$_2$ line to allow their excitation by a laser system discussed below.

\section{Hyperfine structure of H$_2^+$} \label{firstpart}

\subsection{Hyperfine Hamiltonian}

The following notations are used throughout this paper: ${\bf S}_e$ and ${\bf I}_1$, ${\bf I}_2$ are respectively the electron spin and the spins of both protons, with $S_e = I_1 = I_2 = 1/2$. We introduce the total nuclear spin  ${\bf I} = {\bf I}_1 + {\bf I}_2$, where $I$ is equal to 0 or 1. The total orbital angular momentum quantum number is denoted $L$. Note that due to the Pauli symmetrization, and taking into account that the electron is in the ground $1s\sigma_g$ state, the total nuclear spin $I$ is equal to 0 when $L$ is even, and to 1 when $L$ is odd.

The hyperfine effective hamiltonian of the H$_2^+$ molecular ion is taken in a form~\cite{korobov2006}:
\begin{eqnarray}
H_{\rm hfs}&=&b_F({\bf I\cdot S}_e)+c_e({\bf L\cdot S}_e)+c_I({\bf L\cdot I}) \nonumber \\
&+&\frac{d_1}{(2L-1)(2L+3)}\left( \frac{2}{3}{\bf L}^2({\bf I\cdot S}_e)-[({\bf L\cdot I})({\bf L\cdot S}_e)+({\bf L\cdot S}_e)({\bf L\cdot I})]\right)\nonumber\\
&+&\frac{d_2}{(2L-1)(2L+3)}\left(\frac{1}{3}{\bf L}^2{\bf I}^2-\frac{1}{2}({\bf L\cdot I})-({\bf L\cdot I})^2\right). \label{Hhfs}
\end{eqnarray}
The numerical values of the coefficients $b_F,c_e,c_I,d_1,d_2$ have been computed with a relative accuracy of $\mathcal{O}(\alpha^2)$~\cite{korobov2006} using a variational method, for all ro-vibrational levels $(v,L)$ with $0 \leq L \leq 4$ and $0 \leq v \leq 4$.

If $I \neq 0$, the strongest coupling is the spin-spin electron-proton interaction, i.e. the first term in equation~(\ref{Hhfs}). This interaction determines the principal splitting of the ro-vibrational levels of H$_2^+$. With this consideration in mind, the preferable coupling scheme of angular momentum operators is
\begin{equation}
{\bf F} = {\bf S}_e+{\bf I}, \hspace{5mm} {\bf J}={\bf L}+{\bf F}. \label{coupling}
\end{equation}
\begin{table}[b]
\begin{tabular}{|@{\hspace{3mm}}c@{\hspace{3mm}}|@{\hspace{3mm}}c@{\hspace{3mm}}|@{\hspace{3mm}}c@{\hspace{3mm}}|@{\hspace{3mm}}c@{\hspace{3mm}}|@{\hspace{3mm}}c@{\hspace{3mm}}|@{\hspace{3mm}}c@{\hspace{3mm}}|}
\hline
$L$&$S_e$&$I$&$F$&$J$&$n$\\
\hline
\vrule width0pt height11.5pt depth7pt
0&$\frac{1}{2}$&0&$\frac{1}{2}$&$\frac{1}{2}$&1\\
\hline
\vrule width0pt height11.5pt depth0pt
1&$\frac{1}{2}$&1&$\frac{1}{2}$&$\frac{1}{2}$, $\frac{3}{2}$&5\\
\vrule width0pt height12pt depth7pt
 &      &  &$\frac{3}{2}$&$\frac{1}{2}$, $\frac{3}{2}$, $\frac{5}{2}$&\\
\hline
\vrule width0pt height11.5pt depth7pt
even&$\frac{1}{2}$&0&$\frac{1}{2}$&$L\!-\!\frac{1}{2}$, $L\!+\!\frac{1}{2}$&2\\
\hline
\vrule width0pt height11.5pt depth0pt
odd&$\frac{1}{2}$&1&$\frac{1}{2}$&$L\!-\!\frac{1}{2}$, $L\!+\!\frac{1}{2}$&6\\
\vrule width0pt height12pt depth7pt
      &               &  &$\frac{3}{2}$&$L\!-\!\frac{3}{2}$, $L\!-\!\frac{1}{2}$, $L\!+\!\frac{1}{2}$, $L\!+\!\frac{3}{2}$&\\
\hline
\end{tabular}
\caption{\label{FJvalues}Possible values of $F$ and $J$ as a function of $L$. $n$ is the number of hyperfine levels.}
\end{table}
The possible values of $F$ and $J$, as well as the number of hyperfine levels, are given in Table~\ref{FJvalues} for each value of $L$. The hyperfine structure is much simpler for the states of even $L$, where only the value $F = 1/2$ is allowed since the total nuclear spin is zero.

\subsection{Hyperfine states}

In order to obtain the hyperfine eigenstates and frequency shifts, it is necessary to diagonalize the Hamiltonian~(\ref{Hhfs}). This is immediate when $L$ is even: the effective Hamiltonian reduces to $c_e({\bf L\cdot S}_e)$, and can be written
\begin{equation}
H_{\rm hfs}= \frac{c_e}{2}\left({\bf J}^2-{\bf L}^2-{\bf S}_e^2\right)
\end{equation}
Its eigenstates are the states
$|v,L,S_e\!=\!\frac{1}{2},I\!=\!0,F\!=\!\frac{1}{2},J,M_J\rangle$ coupled according to the angular summation scheme~(\ref{coupling}), the corresponding energy shifts are :
\begin{eqnarray}
\left\langle v,L,\frac{1}{2},0,\frac{1}{2},L\!-\!\frac{1}{2}\>\Bigl|\>H_{\rm hfs}\>\Bigr|\>v,L,\frac{1}{2},0,\frac{1}{2},L\!-\!\frac{1}{2}\right\rangle&=&-\frac{L\!+\!1}{2}c_e, \qquad (L \neq 0),\\
\left\langle v,L,\frac{1}{2},0,\frac{1}{2},L\!+\!\frac{1}{2}\>\Bigl|\>H_{\rm hfs}\>|\>v,L,\frac{1}{2},0,\frac{1}{2},L\!+\!\frac{1}{2}\right\rangle&=&\frac{L}{2}c_e.
\end{eqnarray}
\begin{table}
\begin{tabular}{|@{\hspace{3mm}}c@{\hspace{3mm}}|@{\hspace{3mm}}c@{\hspace{3mm}}|@{\hspace{3mm}}c@{\hspace{3mm}}|@{\hspace{3mm}}r@{\hspace{3mm}}|}
\hline
$L$ & $v$ & $J\!=\!L\!-\!1/2$ & $J\!=\!L\!+\!1/2$ \\
\hline
0 & 0 &     & 0.0000 \\
0 & 1 &     & 0.0000 \\
\hline
2 & 0 & $-$63.2438 & 42.1625 \\
2 & 1 & $-$59.3574 & 39.5716 \\
\hline
\end{tabular}
\caption{\label{tab-shiftpair}Hyperfine splitting (in MHz) for the ro-vibrational levels $(v,L)$ with $L = 0,2$ and $v=0,1$. All digits are converged. The relative theoretical accuracy is $\mathcal{O}(\alpha^2)$, which corresponds to an uncertainty of a few kHz.}
\end{table}
All energy shifts for $L\!=\!0,2$ and $v\!=\!0,1$ are given in Table~\ref{tab-shiftpair}. The relative theoretical accuracy is $\mathcal{O}(\alpha^2)$, corresponding to the limit of the Breit-Pauli Hamiltonian used in \cite{korobov2006} to compute the hyperfine coefficients. The numerical accuracy is higher, which is why more digits are given here, as well as in Table~\ref{tab-shiftimpair} below; the extra digits will become useful when higher-order corrections to the hyperfine structure are computed.

\vspace{5mm}

The case of odd $L$ is more complicated. The operators involved in the expression of $H_{\rm hfs}$ are $\mathbf{I\cdot S}_e$, $\mathbf{L\cdot S}_e$, $\mathbf{L\cdot I}$, ${\bf L}^2$ and ${\bf I}^2$. Note that they all commute with $\mathbf{L}^2$, ${\bf S}_e^2$, ${\bf I}^2$, ${\bf J}^2$ and $J_z$, but the terms $\mathbf{L\cdot S}_e$ and $\mathbf{L\cdot I}$ do not commute with $\mathbf{F}^2$. As a consequence, $F$ is an approximate quantum number only. There is a degeneracy in $M_J$, so that it suffices to diagonalize the restriction of $H_{\rm hfs}$ to a subspace of given $M_J$. In the following, $M_J$ is set to $1/2$.

\begin{figure}
\includegraphics[width=8.5cm]{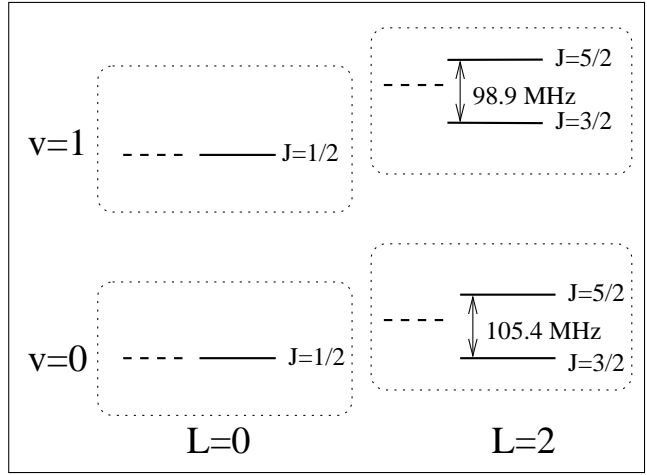}
\caption{\label{hyp-splitpair}Hyperfine splitting of the ro-vibrational levels $(v,L)$ with $L=0,2$ and $v=0,1$. The spacings between hyperfine states are proportional to the frequency difference. That scale is not respected for the rotational and vibrational spacings.}
\end{figure}

\begin{figure}
\includegraphics[width=8.5cm]{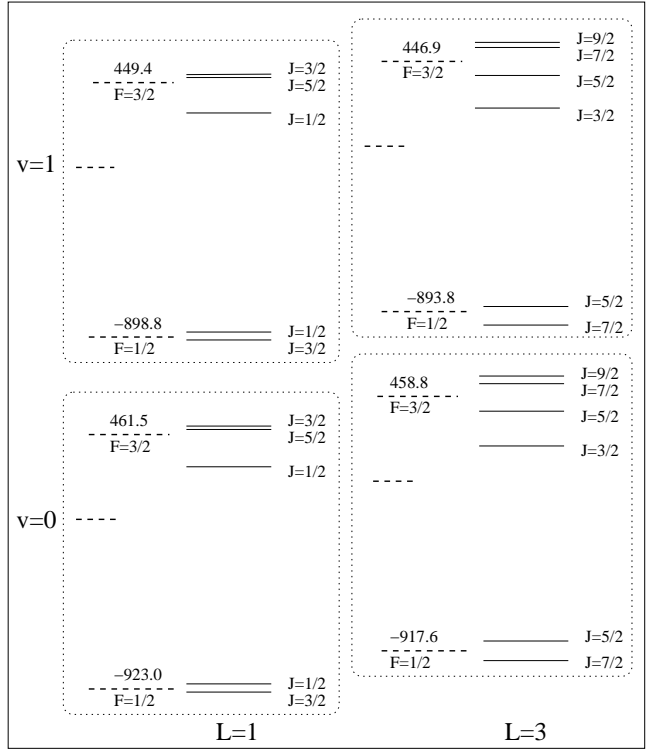}
\caption{\label{hyp-splitimpair}Same as Fig.~\ref{hyp-splitpair}, with $L=1,3$. Within a given $\tilde{F}$ multiplet, the spacings between $J$ states are proportional to the frequency difference. That scale is not respected for the other spacings. The frequency shift of the center of the $\tilde{F}$ multiplets with respect to the spin-independent level, are indicated in MHz.}
\end{figure}

Let us consider a set of states:
\[
\begin{array}{@{}l}
|F\!=\!\frac{3}{2},J\!=\!L\!+\!\frac{3}{2}\rangle,\quad
|F\!=\!\frac{3}{2},J\!=\!L\!+\!\frac{1}{2}\rangle,\quad
|F\!=\!\frac{1}{2},J\!=\!L\!+\!\frac{1}{2}\rangle,
\\[3mm]
|F\!=\!\frac{3}{2},J\!=\!L\!-\!\frac{1}{2}\rangle,\quad
|F\!=\!\frac{1}{2},J\!=\!L\!-\!\frac{1}{2}\rangle,\quad
|F\!=\!\frac{3}{2},J\!=\!L\!-\!\frac{3}{2}\rangle,
\end{array}
\]
where the last {\em ket} exists only if $L\ge3$. We will refer to them as pure states. The matrix representing $H_{\rm hfs}$ in this basis can be derived by use of the following relations
\begin{eqnarray}
\mathbf{I\cdot S}_e &=& \frac{1}{2}\left(\mathbf{F}^2-\mathbf{I}^2-\mathbf{S}_e^2\right)=\frac{1}{2}\left(\mathbf{F}^2-\frac{11}{4}\right) \label{terme1}\\
\left\langle FJ|\mathbf{L\cdot S}_e|F'J\right\rangle &=&
   (-1)^{J+L+F}
   \left\{\begin{array}{ccc}
                  L&1&L\\
                  F'&J&F
          \end{array}
   \right\}
\sqrt{L(L\!+\!1)(2L\!+\!1)}\left\langle S_e,I,F\|S_e\|S_e,I,F'\right\rangle \\
\left\langle FJ|\mathbf{L\cdot I}|F'J\right\rangle &=&
(-1)^{J+L+F}\left\{\begin{array}{ccc}
                      L&1&L\\
                      F'&J&F
                   \end{array}
            \right\}
\sqrt{L(L\!+\!1)(2L\!+\!1)}\left\langle S_e,I,F\|I\|S_e,I,F'\right\rangle \\
{\bf L}^2\>{\bf I}^2 &=& 2L(L+1)\
\end{eqnarray}
and the reduced matrices of ${\bf S}_e$ and ${\bf I}$ on the subspaces $S = \left\{ F = \frac{3}{2}, F = \frac{1}{2} \right\}$ (see Eq.~(91) of Ref.~\cite{messiah})
\begin{equation}
\| {\bf S}_e \| = \left(\begin{array}{cc}
\frac{\sqrt{15}}{3}&-\frac{2}{\sqrt{3}}\\
\frac{2}{\sqrt{3}}&-\frac{\sqrt{6}}{6}
\end{array}\right), \hspace{5mm}
\| {\bf I} \| = \left(\begin{array}{cc}
\frac{2 \sqrt{15}}{3}& \frac{2}{\sqrt{3}}\\
-\frac{2}{\sqrt{3}}&\frac{2 \sqrt{6}}{3}
\end{array}\right). \label{redux}
\end{equation}
Since there is no coupling between different $J$ states, the shape of  $H_{\rm hfs}$ is the following:
\begin{equation}
H_{\rm hfs}=\left(\begin{array}{cccccc}
A&0&0&0&0&0\\
0&B&C&0&0&0\\
0&C&D&0&0&0\\
0&0&0&E&G&0\\
0&0&0&G&H&0\\
0&0&0&0&0&K
\end{array}\right) \label{matrixshape}
\end{equation}
The nonzero coefficients are calculated from equations~(\ref{terme1}-\ref{redux}):
\begin{eqnarray}
A &=& \frac{b_F}{2} + \frac{L}{2} \left( c_e + 2 c_I - \frac{1}{3} \frac{2 d_1 + d_2}{2L+3} \right)\\
B &=& \frac{b_F}{2} + \frac{L-3}{6} \left( c_e + 2 c_I \right) + \frac{L+3}{6} \frac{2 d_1 + d_2}{2L+3} \\
C &=& \frac{\sqrt{L(2L+3)}}{3} \left( c_e - c_I \right) - \frac{\sqrt{L}}{6 \sqrt{2L+3}} \left( d_1 - d_2 \right) \\
D &=& -b_F - \frac{L}{6} \left( c_e - 4 c_I \right) \\
E &=& \frac{b_F}{2} - \frac{L+4}{6} \left( c_e + 2 c_I \right) + \frac{L-2}{6} \frac{2 d_1 + d_2}{2L-1} \\
G &=& \frac{\sqrt{(L+1)(2L-1)}}{3} \left( c_e - c_I \right) + \frac{\sqrt{L+1}}{6\sqrt{2L-1}} \left( d_1 - d_2 \right) \\
H &=& -b_F + \frac{L+1}{6} \left( c_e - 4 c_I \right) \\
K &=& \frac{b_F}{2} - \frac{L+1}{2} \left( c_e + 2 c_I + \frac{1}{3} \frac{2 d_1 + d_2}{2L-1} \right)
\end{eqnarray}
\begin{table}
\begin{tabular}{|@{\hspace{3mm}}c@{\hspace{3mm}}|@{\hspace{3mm}}c@{\hspace{3mm}}|@{\hspace{3mm}}c@{\hspace{3mm}}|@{\hspace{3mm}}c@{\hspace{3mm}}|@{\hspace{3mm}}r@{\hspace{3mm}}|@{\hspace{3mm}}c@{\hspace{3mm}}|}
\hline
$L$& $v$ & $\tilde F$ & $J$ & $\Delta E_{\rm hfs}$ & $\left[ C^{\pm}_1,C^{\pm}_3 \right]$ \\
\hline
 &   & 3/2 & 5/2 &    474.1063 & $[0,1]$ \\
 &   & 3/2 & 3/2 &    481.9534 & $[0.015612,0.999878]$ \\
1& 0 & 1/2 & 3/2 & $-$930.4332 & $[-0.999878,0.015612]$ \\
 &   & 3/2 & 1/2 &    385.3985 & $[0.038891,0.999243]$ \\
 &   & 1/2 & 1/2 & $-$910.7579 & $[-0.999243,0.038891]$ \\
\hline
 &   & 3/2 & 5/2 &    461.2574 & $[0,1]$ \\
 &   & 3/2 & 3/2 &    468.5247 & $[0.015074,0.999886]$ \\
1& 1 & 1/2 & 3/2 & $-$905.7836 & $[-0.999886,0.015074]$ \\
 &   & 3/2 & 1/2 &    377.9948 & $[0.037345,0.999302]$ \\
 &   & 1/2 & 1/2 & $-$887.2491 & $[-0.999302,0.037345]$\\
\hline
  &  & 3/2 & 9/2 &    507.2568 & $[0,1]$ \\
  &  & 3/2 & 7/2 &    489.5257 & $[0.042115,0.999113]$ \\
3 &0 & 1/2 & 7/2 & $-$941.1034 & $[-0.999113,0.042115]$ \\
  &  & 3/2 & 5/2 &    423.6342 & $[0.061812,0.998088]$ \\
  &  & 1/2 & 5/2 & $-$894.6614 & $[-0.998088,0.061812]$ \\
  &  & 3/2 & 3/2 &    341.5540 & $[0,1]$\\
\hline
 &   & 3/2 & 9/2 &    492.3817 & $[0,1]$ \\
 &   & 3/2 & 7/2 &    475.5771 & $[0.040656,0.999173]$ \\
3& 1 & 1/2 & 7/2 & $-$915.7408 & $[-0.999173,0.040656]$ \\
 &   & 3/2 & 5/2 &    413.6810 & $[0.059441,0.998232]$ \\
 &   & 1/2 & 5/2 & $-$872.0486 & $[-0.998232,0.059441]$ \\
 &   & 3/2 & 3/2 &    336.9246 & $[0,1]$\\
\hline
\end{tabular}
\caption{\label{tab-shiftimpair}Hyperfine splitting (in MHz) and eigenstates for the ro-vibrational levels $(v,L)$ with $L\!=\!1,3$ and $v\!=\!0,1$. All digits are converged. The relative theoretical accuracy on the frequency shifts, as well as on the smaller of the two coefficients $\left[ C^{\pm}_1,C^{\pm}_3 \right]$, is $\mathcal{O}(\alpha^2)$. This corresponds to a few tens of kHz for the frequency shifts.}
\end{table}
The eigenstates of $J\!=\!L \pm \frac{3}{2}$ are pure states of angular coupling: $|v,L,S_e\!=\!1/2,I\!=\!1,F\!=\!3/2,J\!=\!L \pm 3/2\rangle$, while the eigenstates of $J\!=\!L \pm \frac{1}{2}$ are linear combinations of $F\!=\!1/2$ and $F\!=\!3/2$ states, obtained by diagonalization of the $2\times2$ sub-matrices appearing in~(\ref{matrixshape}):
\begin{equation}
\textstyle
|\,v,L,S_e,I,\tilde{F}, J\!=\!L\pm\frac{1}{2} \rangle \equiv
C^{\pm}_1 \,|\,v,L,\frac{1}{2},1,\frac{1}{2},L \pm \frac{1}{2}\rangle
\>+\>
C^{\pm}_3 \,|\,v,L,\frac{1}{2},1,\frac{3}{2},L \pm \frac{1}{2}\rangle. \label{eq-coefCi1}
\end{equation}
We will refer to them as mixed states. The coefficients $C^{\pm}_1$ and $C^{\pm}_3$ are calculated in Table~\ref{tab-shiftimpair} together with the hyperfine frequency shifts, for $L\!=\!1,3$ and $v\!=\!0,1$. The mixing between $F\!=\!1/2$ and $F\!=\!3/2$ states is weak, so that the states can be labeled by the dominant $F$, noted as $\tilde{F}$. The hyperfine splitting of the first ro-vibrational levels \cite{suppl} is shown in Figs.~\ref{hyp-splitpair} and~\ref{hyp-splitimpair}.
%
%%%%%%%%%%%%%%%%%%%%%%%%%%%%%%%%%%%%%%%%%%%%%%%%%%%%%%%%%%%%%%%%%%%%%%%%%%%%
\section{Two-photon transitions} \label{secondpart}

\subsection{Two-photon transition operator}

In this paragraph, we present the general theory of two-photon transitions with arbitrary excitation polarizations, as developed by G. Grynberg in~\cite{ThGrynberg}. Let us consider an H$_2^+$ ion irradiated by two beams of polarizations ${\bf \epsilon}_1$ and ${\bf \epsilon}_2$. The transition probability between two states $|\phi\rangle$ and $|\psi\rangle$ by absorption of one photon in each wave is proportional to
\begin{equation}
| \langle \phi|^SQ_{{\bf \epsilon_1\epsilon_2}}|\psi\rangle|^2
\end{equation}
where
\begin{equation}
^SQ_{{\bf \epsilon_1\epsilon_2}}=\frac{1}{2}\left(Q_{{\bf \epsilon_1\epsilon_2}}+Q_{{\bf \epsilon_2\epsilon_1}}\right)
\end{equation}
is the two-photon transition operator, with
\begin{equation}
Q_{{\bf \epsilon_1\epsilon_2}}=\mathbf{d}\cdot\
%boldsymbol{
\epsilon_1
%}\>
\frac{1}{H-E}
%\>
\mathbf{d}\cdot
%\boldsymbol{
\epsilon_2
%}
\end{equation}
In this expression, ${\bf d}$ is the dipole operator, $H$ is the full hamiltonian and $E$ the intermediate state energy. If the excitation polarizations are chosen among the standard polarizations $\pi$, $\sigma^+$ and $\sigma^-$, the two-photon transition operator reads
\begin{equation}
^SQ_{q_1 q_2}=\frac{1}{2}\left(Q_{q_1 q_2}+Q_{q_2 q_1}\right), \hspace{5mm} Q_{q_1 q_2}=d_{q_1}\frac{1}{H-E}d_{q_2}
\end{equation}
where $d_{q_i}$ ($q_i=-1,0,1$) are the standard components of ${\bf d}$. Tensor $Q_{q_1 q_2}$ has a rank 2 and can be represented in terms of irreducible tensors:
\begin{equation}
Q^{(k)}_q=\sum_{q_1,q_2}\bigl\langle kq\big|11q_1q_2\bigr\rangle\; Q_{q_1 q_2}, \hspace{5mm} k=0,1,2. \label{eqQkq}
\end{equation}
Inverting this expression, one finds:
\begin{equation}
^SQ_{q_1 q_2}=\sum_{q=-2}^{2} a^{(2)}_q\  Q^{(2)}_q + a^{(0)}_0 Q^{(0)}_0,\label{eqQa}
\end{equation}
where
\begin{equation}
a^{(k)}_q=\bigl\langle11q_1q_2\big|kq\bigr\rangle. \label{eqakq}
\end{equation}
Table~\ref{table_akq} gives the values of the coefficients $a^{(k)}_q$ for all combinations of the standard polarizations.
\begin{table}
\begin{tabular}{|@{\hspace{3mm}}c@{\hspace{3mm}}|@{\hspace{3mm}}c@{\hspace{3mm}}|@{\hspace{3mm}}c@{\hspace{3mm}}|@{\hspace{3mm}}c@{\hspace{3mm}}|}
\hline
&$\sigma_-$&$\pi$&$\sigma_+$\\[-1mm]
\vrule width0pt height 10pt depth5pt
&$q_1\!=\!-1$&$q_1\!=\!0$&$q_1\!=\!1$\\
\hline
\vrule width0pt height 12pt depth0pt
$\sigma_-$&$a^{(2)}_q=\delta_{q,-2}$ &$a^{(2)}_q=\frac{\sqrt{2}}{2}\delta_{q,-1}$&$a^{(2)}_q=\frac{\sqrt{6}}{6}\delta_{q,0}$\\
\vrule width0pt height 10pt depth6pt
$q_2\!=\!-1$&$a^{(0)}_0=0$&$a^{(0)}_0=0$&$a^{(0)}_0=\frac{\sqrt{3}}{3}$\\
\hline
\vrule width0pt height 12pt depth0pt
$\pi$&$a^{(2)}_q=\frac{\sqrt{2}}{2}\delta_{q,-1}$ &$a^{(2)}_q=\sqrt{\frac{2}{3}}\delta_{q,0}$&$a^{(2)}_q=\frac{\sqrt{2}}{2}\delta_{q,1}$ \\
\vrule width0pt height 10pt depth6pt
$q_2\!=\!0$&$a^{(0)}_0=0$&$a^{(0)}_0=-\frac{\sqrt{3}}{3}$&$a^{(0)}_0=0$   \\
\hline
\vrule width0pt height 12pt depth0pt
$\sigma_+$&$a^{(2)}_q=\frac{\sqrt{6}}{6}\delta_{q,0}$&$a^{(2)}_q=\frac{\sqrt{2}}{2}\delta_{q,1}$&$a^{(2)}_q=\delta_{q,2}$\ \\
\vrule width0pt height 10pt depth6pt
$q_2\!=\!1$&$a^{(0)}_0=\frac{\sqrt{3}}{3}$&$a^{(0)}_0=0$&$a^{(0)}_0=0$ \\ \hline
\end{tabular}
\caption{\label{table_akq}Values of the coefficients $a^{(k)}_q$ for all combinations of the standard polarizations.}
\end{table}
\subsection{Two-photon matrix elements between hyperfine levels}
We consider a two-photon transition between the hyperfine states $|\phi\rangle\!=\!|v,L,S_e,I,F,J,M_J\rangle\!=\!|g,J,M_J\rangle$ and $|\psi\rangle=|v',L',S_e,I',F',J',M_J'\rangle=|e,J',M_J'\rangle$ with standard excitation polarizations $q_1$, $q_2$. In order to simplify the expressions, we restrict the presentation to a case where the initial and final states are pure states; the results will be generalized at the end of the paragraph. Using the Wigner-Eckart theorem, the two-photon matrix element between $|\phi\rangle$ and $|\psi\rangle$ may be expressed as:
\begin{equation}
\left\langle\phi|^SQ_{q_1 q_2}|\psi\right\rangle =
   \sum_{k} a^{(k)}_q \left\langle J'kM_J'q|JM_J\right\rangle
   \frac{\left\langle gJ\|Q^{(k)}\|eJ'\right\rangle}{\sqrt{2J+1}},
\qquad q=q_1+q_2.
\label{eqQk}
\end{equation}
The states $|g,J,M_J\rangle$ and $|e,J',M_J'\rangle$ are degenerate in $M_J$ or $M_J'$. If we assume the initial level to be unpolarized, the transition probability averaged over $M_J$ and $M_J'$ is proportional to the averaged squared matrix element
\begin{equation}
\left[ ^S\bar{Q}_{q_1,q_2}(gJ \rightarrow eJ') \right]^2=\frac{1}{2J+1}\sum_{M_J,M_J'}\left|\left\langle g,J,M_J|^SQ_{q_1q_2}|e,J',M_J'\right\rangle\right|^2.
\end{equation}
Using the orthogonality relations of Clebsch-Gordan coefficients~\cite{messiah}, one obtains~\cite{ThGrynberg}
\begin{equation}
\left[ ^S\bar{Q}_{q_1,q_2}(gJ \rightarrow eJ') \right]^2 =
\frac{1}{2J+1}\sum_{k=0,2}
\frac{\left|a^{(k)}_q\left\langle gJ\|Q^{(k)}\|eJ'\right\rangle\right|^2}{2k+1}
\label{eqSQ2cb}
\end{equation}
The dipole operator and hence $Q^{(k)}$ acts on the spatial variables only. Using Eq.~(89) of Ref.~\cite{messiah} one can write the reduced matrix elements of $Q^{(k)}$ using only orbital wave functions:
\begin{equation}
\left\langle gJ\|Q^{(k)}\|eJ'\right\rangle=
\delta_{I,I'}\delta_{F,F'}(-1)^{J'\!+\!L\!+\!F\!+\!k}\sqrt{2J\!+\!1}\sqrt{2J'\!+\!1}
\left\{\begin{array}{ccc}
          L&k&L'\\
          J'&F&J
       \end{array}\right\}
\left\langle vL\|Q^{(k)}\|v'L'\right\rangle. \label{eq6j}
\end{equation}
In the case where the initial and final states are not pure basis states (i.e. for odd $L$ and $J\!=\!L \pm 1/2$), they can be written according to Eq.~(\ref{eq-coefCi1}):
\begin{equation}
|\tilde g,J\rangle=\sum_{F_i=\frac{1}{2},\frac{3}{2}}\>C_{F_i}\>
|v,L,1/2,1,F_i,J\rangle
\end{equation}

This expression can also be applied to pure states, where one coefficient is equal to zero and the other is equal to one.  It is then straightforward to generalize Eq.~(\ref{eq6j}):
\begin{equation}
\left\langle\tilde gJ\|Q^{(k)}\|\tilde eJ'\right\rangle=
\delta_{I,I'}\sum_{F_i, F'_j}\delta_{F_i,F_j'}C_{F_i}C_{F'_j}(-1)^{J'\!+\!L\!+\!F_i\!+\!k}\sqrt{2J\!+\!1}\sqrt{2J'\!+\!1}
\left\{\begin{array}{ccc}
          L&k&L'\\
          J'&F_i&J
       \end{array}\right\}
\left\langle vL\|Q^{(k)}\|v'L'\right\rangle. \label{eq6jgeneral}
\end{equation}

\subsection{Selection rules}
Since the two-photon transition operator is a sum of operators of rank 0 and 2, the states $|\phi\rangle\!=\!|v,L,S_e,I,F,J,M_J\rangle$ and $|\psi\rangle\!=\!|v',L',S_e,I',F',J',M_J'\rangle$ can be coupled only if $|L-L'|\le 2$ and $|J-J'|\le 2$.

For the ro-vibrational states of H$_2^+$, the total nuclear spin is $I=0$ when $L$ is even and $I=1$ when $L$ is odd. The two-photon transition operator acts on the orbital variables only, which explains the $\delta_{II'}$ factor in Eq.~(\ref{eq6j}) and gives the selection rule $\Delta L\!=\!0$, or $\pm2$.

For the same reason, if we consider pure states, we get the selection rule $\Delta F \!=\! 0$, as can be seen from the $\delta_{FF'}$ factor in Eq.~(\ref{eq6j}). However, due to the mixing between $F\!=\!1/2$ and $F \!=\! 3/2$, transitions between mixed states of different $\tilde{F}$ are weakly allowed.

There are additional selection rules on $M_J$ and $M_J'$ depending on the beam polarizations as can be seen in Table~\ref{table_akq}. A difference $M_J'\!-\!M_J$ must be equal to 0 for $\pi \pi$ and $\sigma_+ \sigma_-$, $+1$ ($-1$) for $\pi \sigma_+$ ($\pi \sigma_-$) and $+2$ ($-2$) for $\sigma_+ \sigma_+$ ($\sigma_- \sigma_-$).

\subsection{Reduced orbital two-photon matrix elements}
The last step consists in the numerical computation of the reduced matrix elements $\left\langle vL\|Q^{(k)}\|v'L'\right\rangle$. This is achieved using the variational approach outlined in Ref.~\cite{korobov2006}. Briefly, the wave function for a state with a total orbital angular momentum $L$ and of a total spatial parity $\pi=(-1)^L$ is expanded as follows:
\begin{equation}\label{exp_main}
\begin{array}{@{}l}
\displaystyle \Psi_{LM}^\pi(\mathbf{R},\mathbf{r}_1) =
       \sum_{l_1+l_2=L}
         \mathcal{Y}^{l_1l_2}_{LM}(\hat{\mathbf{R}},\hat{\mathbf{r}}_1)
         G^{L\pi}_{l_1l_2}(R,r_1,r_2),
\\[4mm]\displaystyle
G_{l_1l_2}^{L\pi}(R,r_1,r_2) = \sum_{n=1}^N \Big\{C_n\,\mbox{Re} \bigl[e^{-\alpha_n R-\beta_n r_1-\gamma_n r_2}\bigr]
%\\[2mm]\displaystyle\hspace{30mm}
+D_n\,\mbox{Im} \bigl[e^{-\alpha_n R-\beta_n r_1-\gamma_n r_2}\bigr] \Big\}.
\end{array}
\end{equation}
where the complex exponents $\alpha$, $\beta$, $\gamma$, are generated in a pseudorandom way. The use of complex exponents instead of real ones allows to reproduce the oscillatory behavior of the vibrational part of the wave function and improves the convergence rate. Since very high accuracy is not required for transition probabilities, relatively small basis lengths of $N = 700-1000$ were used, providing a relative accuracy of a few parts in $10^{9}$ for the nonrelativistic energies, and a few parts in $10^{5}$ for the matrix elements.

The reduced matrix elements $\left\langle vL\|Q^{(k)}\|v'L'\right\rangle$ are divided into three terms corresponding to the possible values $L\!-\!1$, $L\!+\!1$, $L$ for the angular momentum of the intermediate state. The three following terms are evaluated numerically:
\begin{eqnarray}
a_-&=&-\sum_{v''}\frac{\left\langle vL\|d\|v''L\!-\!1\right\rangle
                       \left\langle v''L\!-\!1\|d\|v'L'\right\rangle}
                          {\sqrt{(2L\!+\!1)(2L'\!+\!1)}(\omega-E_{v''L\!-\!1})}\\
a_+&=&-\sum_{v''}\frac{\left\langle vL\|d\|v''L\!+\!1\right\rangle
                       \left\langle v''L\!+\!1||d||v'L'\right\rangle}
                       {\sqrt{(2L\!+\!1)(2L'\!+\!1)}(\omega-E_{v''L\!+\!1})}\\
a_0&=&\sum_{v''}\frac{\left\langle vL\|d\|v''L\right\rangle
                      \left\langle v''L\|d\|v'L'\right\rangle}
                      {\sqrt{(2L\!+\!1)(2L'\!+\!1)}(\omega-E_{v''L})}
\end{eqnarray}
where $E_{v'',L''}$ is the energy of the intermediate state $|v''L''\rangle$ and $\omega=(E_{v'L'}-E_{vL})/2$ is the photon energy. The reduced matrix elements of $Q^{(k)}$ are related to $a_-$, $a_+$, $a_0$ in the following way:
\begin{eqnarray}
\frac{\left\langle vL\|Q^{(0)}\|v'L\right\rangle}{\sqrt{2L+1}}&=&-\frac{\sqrt{3}}{3}\>\bigl(a_-+a_0+a_+\bigr)\>,\\
\frac{\left\langle vL\|Q^{(2)}\|v'L\!-\!2\right\rangle}{\sqrt{2L+1}}&=&-\sqrt{\frac{2L\!-\!3}{2L\!-\!1}}\>a_-\>,\\
\frac{\left\langle vL\|Q^{(2)}\|v'L\right\rangle}{\sqrt{2L+1}}&=&-\frac{1}{\sqrt{6}}\sqrt{(2L\!+\!3)(2L\!-\!1)L(L\!+\!1)}\>\left[\frac{a_-}{L(2L\!-\!1)}-\frac{a_0}{L(L\!+\!1)}+\frac{a_+}{(2L\!+\!3)(L\!+\!1)}\right]\>,\\
\frac{\left\langle vL\|Q^{(2)}\|v'L\!+\!2\right\rangle}{\sqrt{2L+1}}&=&-\sqrt{\frac{2L\!+\!5}{2L\!+\!3}}\;a_+\>.
\end{eqnarray}
The reduced matrix elements of $Q^{(0)}$ and $Q^{(2)}$ for the transitions $(v\!=\!0,L) \rightarrow (v'\!=\!1,L)$ with $0 \leq L \leq 3$, are given in Table~\ref{Q0Q2}.
\begin{table}
\begin{tabular}{|@{\hspace{3mm}}c@{\hspace{3mm}}|@{\hspace{3mm}}c@{\hspace{3mm}}|@{\hspace{3mm}}c@{\hspace{3mm}}|}
\hline
$L$&$\left\langle0L\|Q^{(0)}\|1L\right\rangle$&$\left\langle0L\|Q^{(2)}\|1L\right\rangle$\\
\hline
0&0.7255&0\\
1&1.261&0.7753\\
2&1.640&0.8541\\
3&1.962&0.9903\\
\hline
\end{tabular}
\caption{\label{Q0Q2}Reduced matrix elements of the operators $Q^{(0)}$ and $Q^{(2)}$ for the transitions $(v\!=\!0,L) \rightarrow (v'\!=\!1,L)$ with $0 \leq L \leq 3$, in atomic units.}
\end{table}
\subsection{Two-photon transition spectra}
%
%%%%%%%%%%%%%%%%%%%%%%%%%%%%%%%%%%%%%%%%%%%%%%%%%%%%%%%%%%%%%%%%%%%
\begin{table}
\begin{tabular}{|@{\hspace{3mm}}c@{\hspace{3mm}}|@{\hspace{3mm}}c@{\hspace{3mm}}|@{\hspace{3mm}}c@{\hspace{3mm}}|@{\hspace{3mm}}c@{\hspace{3mm}}|@{\hspace{3mm}}c@{\hspace{3mm}}|@{\hspace{3mm}}c@{\hspace{3mm}}|@{\hspace{3mm}}c@{\hspace{3mm}}|}
\hline
$L$ & $\Delta f$  &  $J$  &  $J'$  & $\pi \pi$ & $\sigma^+ \sigma^+$ & $\sigma^+ \sigma^-$ \\
\hline
0 &    0.0000  &  1/2 & 1/2 &     0.1754 &      0.0000 &      0.1754  \\
\hline
  & $-$50.7600 & 5/2 & 3/2  &     0.0039 &      0.0058 &      0.0010  \\
2 & $-$1.2955  & 5/2 & 5/2  &     0.1949 &      0.0233 &      0.1832  \\
  &    1.9432  & 3/2 & 3/2  &     0.1929 &      0.0204 &      0.1827  \\
  &    51.4077 & 3/2 & 5/2  &     0.0058 &      0.0088 &      0.0015  \\
\hline
\end{tabular}
\caption{\label{tab-probapair}Average two-photon matrix elements $\left[ ^S\bar{Q}_{q_1,q_2} \right]^2$ given by Eq.~(\ref{eqSQ2cb}) between the ro-vibrational levels $(v=0,L)$ and $(v=1,L)$ with $L=0,2$, in atomic units. $\Delta f$ is the hyperfine shift of the transition frequency in MHz.}
\end{table}
\begin{table}
\tiny
\begin{tabular}{|@{\hspace{3mm}}c@{\hspace{3mm}}|@{\hspace{3mm}}c@{\hspace{3mm}}|@{\hspace{3mm}}c@{\hspace{3mm}}|@{\hspace{3mm}}c@{\hspace{3mm}}|@{\hspace{3mm}}c@{\hspace{3mm}}|@{\hspace{3mm}}c@{\hspace{3mm}}|}
\hline
$\Delta f$  & $(F,J)$ & $(F',J')$ & $\pi \pi$  & $\sigma^+ \sigma^+$ & $\sigma^+ \sigma^-$ \\
\hline
$-$693.869 & (3/2,3/2) & (1/2,3/2) & 1.028e$-$05 & 1.534e$-$05 & 2.607e$-$06 \\
$-$689.945 & (3/2,5/2) & (1/2,3/2) & 2.550e$-$06 & 3.824e$-$06 & 6.374e$-$07 \\
$-$684.601 & (3/2,3/2) & (1/2,1/2) & 1.003e$-$05 & 1.505e$-$05 & 2.509e$-$06 \\
$-$680.678 & (3/2,5/2) & (1/2,1/2) & 1.118e$-$05 & 1.677e$-$05 & 2.794e$-$06 \\
$-$645.591 & (3/2,1/2) & (1/2,3/2) & 5.090e$-$05 & 7.635e$-$05 & 1.273e$-$05 \\
$-$636.324 & (3/2,1/2) & (1/2,1/2) & 4.229e$-$07 & 0.000e$-$01 & 4.229e$-$07 \\
\hline
$-$51.979 & (3/2,3/2) & (3/2,1/2) & 1.329e$-$03 & 1.993e$-$03 & 3.322e$-$04 \\
$-$48.056 & (3/2,5/2) & (3/2,1/2) & 8.003e$-$03 & 1.201e$-$02 & 2.001e$-$03 \\
$-$10.348 & (3/2,3/2) & (3/2,5/2) & 1.683e$-$02 & 2.524e$-$02 & 4.207e$-$03 \\
 $-$6.714 & (3/2,3/2) & (3/2,3/2) & 1.853e$-$01 & 1.281e$-$02 & 1.789e$-$01 \\
 $-$6.424 & (3/2,5/2) & (3/2,5/2) & 1.842e$-$01 & 1.122e$-$02 & 1.786e$-$01 \\
 $-$3.702 & (3/2,1/2) & (3/2,1/2) & 1.767e$-$01 & 0.000e$-$01 & 1.767e$-$01 \\
 $-$2.791 & (3/2,5/2) & (3/2,3/2) & 1.122e$-$02 & 1.683e$-$02 & 2.804e$-$03 \\
    2.487 & (1/2,1/2) & (1/2,3/2) & 2.666e$-$02 & 3.999e$-$02 & 6.665e$-$03 \\
   11.754 & (1/2,1/2) & (1/2,1/2) & 1.767e$-$01 & 0.000e$-$01 & 1.767e$-$01 \\
   12.325 & (1/2,3/2) & (1/2,3/2) & 1.901e$-$01 & 2.002e$-$02 & 1.801e$-$01 \\
   21.592 & (1/2,3/2) & (1/2,1/2) & 1.333e$-$02 & 2.000e$-$02 & 3.333e$-$03 \\
   37.929 & (3/2,1/2) & (3/2,5/2) & 2.401e$-$02 & 3.601e$-$02 & 6.002e$-$03 \\
   41.563 & (3/2,1/2) & (3/2,3/2) & 2.657e$-$03 & 3.986e$-$03 & 6.643e$-$04 \\
\hline
644.376 & (1/2,1/2) & (3/2,1/2) & 4.229e$-$07 & 0.000e$-$01 & 4.229e$-$07 \\
654.214 & (1/2,3/2) & (3/2,1/2) & 2.387e$-$05 & 3.581e$-$05 & 5.968e$-$06 \\
686.008 & (1/2,1/2) & (3/2,5/2) & 3.637e$-$05 & 5.455e$-$05 & 9.092e$-$06 \\
689.641 & (1/2,1/2) & (3/2,3/2) & 2.000e$-$05 & 3.000e$-$05 & 4.999e$-$06 \\
695.845 & (1/2,3/2) & (3/2,5/2) & 4.102e$-$06 & 6.153e$-$06 & 1.026e$-$06 \\
699.479 & (1/2,3/2) & (3/2,3/2) & 1.020e$-$05 & 1.522e$-$05 & 2.588e$-$06 \\
\hline
\end{tabular}
\caption{\label{tab-probaL1} Same as Table~\ref{tab-probapair}, with $L=1$.}
\end{table}
\begin{table}
\tiny
\begin{tabular}{|@{\hspace{3mm}}c@{\hspace{3mm}}|@{\hspace{3mm}}c@{\hspace{3mm}}|@{\hspace{3mm}}c@{\hspace{3mm}}|@{\hspace{3mm}}c@{\hspace{3mm}}|@{\hspace{3mm}}c@{\hspace{3mm}}|@{\hspace{3mm}}c@{\hspace{3mm}}|}
\hline
$\Delta f$  & $(F,J)$ & $(F',J')$ & $\pi \pi$  & $\sigma^+ \sigma^+$ & $\sigma^+ \sigma^-$ \\
\hline
$-$711.499 & (3/2,9/2) & (1/2,7/2) & 6.739e$-$06 & 1.011e$-$05 & 1.685e$-$06 \\
$-$702.633 & (3/2,7/2) & (1/2,7/2) & 4.143e$-$06 & 5.629e$-$06 & 1.329e$-$06 \\
$-$689.653 & (3/2,9/2) & (1/2,5/2) & 1.179e$-$06 & 1.768e$-$06 & 2.947e$-$07 \\
$-$680.787 & (3/2,7/2) & (1/2,5/2) & 6.647e$-$06 & 9.971e$-$06 & 1.662e$-$06 \\
$-$669.687 & (3/2,5/2) & (1/2,7/2) & 1.073e$-$07 & 1.610e$-$07 & 2.684e$-$08 \\
$-$647.841 & (3/2,5/2) & (1/2,5/2) & 1.457e$-$05 & 2.030e$-$05 & 4.417e$-$06 \\
$-$628.647 & (3/2,3/2) & (1/2,7/2) & 1.764e$-$06 & 2.647e$-$06 & 4.411e$-$07 \\
$-$606.801 & (3/2,3/2) & (1/2,5/2) & 3.055e$-$05 & 4.582e$-$05 & 7.637e$-$06 \\
\hline
$-$85.166 & (3/2,9/2) & (3/2,3/2) & 0.000e$-$01 & 0.000e$-$01 & 0.000e$-$01 \\
$-$76.301 & (3/2,7/2) & (3/2,3/2) & 5.328e$-$04 & 7.992e$-$04 & 1.332e$-$04 \\
$-$46.788 & (3/2,9/2) & (3/2,5/2) & 3.324e$-$04 & 4.986e$-$04 & 8.310e$-$05 \\
$-$43.355 & (3/2,5/2) & (3/2,3/2) & 5.742e$-$03 & 8.613e$-$03 & 1.436e$-$03 \\
$-$37.922 & (3/2,7/2) & (3/2,5/2) & 5.624e$-$03 & 8.437e$-$03 & 1.406e$-$03 \\
$-$15.840 & (3/2,9/2) & (3/2,7/2) & 4.070e$-$03 & 6.106e$-$03 & 1.018e$-$03 \\
$-$10.540 & (1/2,5/2) & (1/2,7/2) & 2.677e$-$03 & 4.015e$-$03 & 6.691e$-$04 \\
 $-$7.438 & (3/2,9/2) & (3/2,9/2) & 1.975e$-$01 & 2.140e$-$02 & 1.868e$-$01 \\
 $-$6.974 & (3/2,7/2) & (3/2,7/2) & 1.906e$-$01 & 1.114e$-$02 & 1.851e$-$01 \\
 $-$4.977 & (3/2,5/2) & (3/2,5/2) & 1.881e$-$01 & 7.309e$-$03 & 1.844e$-$01 \\
 $-$2.315 & (3/2,3/2) & (3/2,3/2) & 1.922e$-$01 & 1.345e$-$02 & 1.855e$-$01 \\
    1.428 & (3/2,7/2) & (3/2,9/2) & 5.087e$-$03 & 7.631e$-$03 & 1.272e$-$03 \\
   11.306 & (1/2,5/2) & (1/2,5/2) & 1.992e$-$01 & 2.394e$-$02 & 1.872e$-$01 \\
   12.681 & (1/2,7/2) & (1/2,7/2) & 1.999e$-$01 & 2.499e$-$02 & 1.874e$-$01 \\
   25.971 & (3/2,5/2) & (3/2,7/2) & 7.498e$-$03 & 1.125e$-$02 & 1.875e$-$03 \\
   34.374 & (3/2,5/2) & (3/2,9/2) & 5.538e$-$04 & 8.308e$-$04 & 1.385e$-$04 \\
   34.527 & (1/2,7/2) & (1/2,5/2) & 2.008e$-$03 & 3.012e$-$03 & 5.019e$-$04 \\
   36.063 & (3/2,3/2) & (3/2,5/2) & 8.616e$-$03 & 1.292e$-$02 & 2.154e$-$03 \\
   67.012 & (3/2,3/2) & (3/2,7/2) & 1.066e$-$03 & 1.599e$-$03 & 2.664e$-$04 \\
   75.414 & (3/2,3/2) & (3/2,9/2) & 0.000e$-$01 & 0.000e$-$01 & 0.000e$-$01 \\
\hline
615.793 & (1/2,5/2) & (3/2,3/2) & 2.202e$-$05 & 3.304e$-$05 & 5.506e$-$06 \\
639.014 & (1/2,7/2) & (3/2,3/2) & 9.467e$-$07 & 1.420e$-$06 & 2.367e$-$07 \\
654.171 & (1/2,5/2) & (3/2,5/2) & 1.136e$-$05 & 1.548e$-$05 & 3.614e$-$06 \\
677.392 & (1/2,7/2) & (3/2,5/2) & 2.496e$-$07 & 3.745e$-$07 & 6.241e$-$08 \\
685.119 & (1/2,5/2) & (3/2,7/2) & 1.062e$-$05 & 1.593e$-$05 & 2.654e$-$06 \\
693.522 & (1/2,5/2) & (3/2,9/2) & 2.124e$-$06 & 3.186e$-$06 & 5.310e$-$07 \\
708.340 & (1/2,7/2) & (3/2,7/2) & 3.025e$-$06 & 3.951e$-$06 & 1.049e$-$06 \\
716.743 & (1/2,7/2) & (3/2,9/2) & 9.039e$-$06 & 1.356e$-$05 & 2.260e$-$06 \\
\hline
\end{tabular}
\caption{\label{tab-probaL3} Same as Table~\ref{tab-probapair}, with $L=3$.}
\end{table}
%%%%%%%%%%%%%%%%%%%%%%%%%%%%%%%%%%%%%%%%%%%%%%%%%%%%%%%%%%%%%%%%%%%%%%%%%
\begin{figure}
\includegraphics[width=8cm]{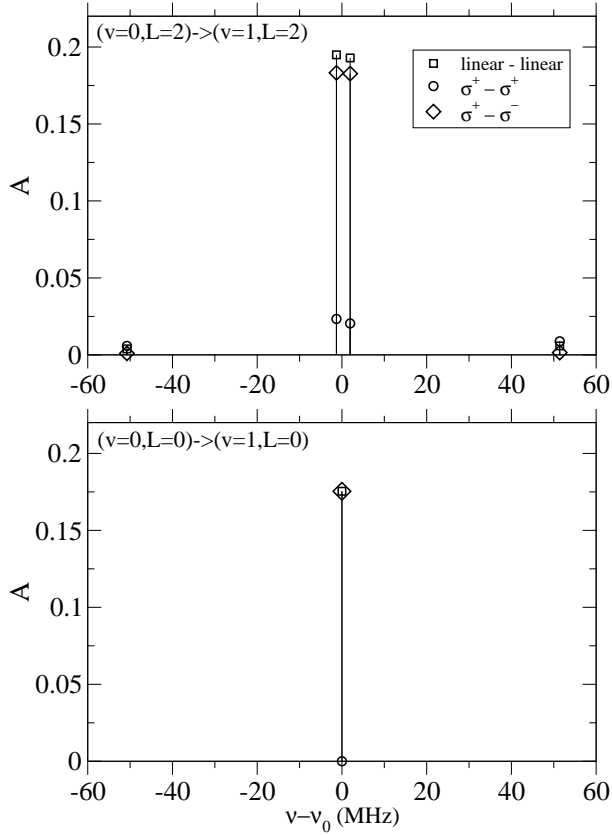}
\caption{\label{fig-probapair}Averaged two-photon matrix elements $A = \left[ ^S\bar{Q}_{q_1,q_2} \right]^2$ in atomic units between the ro-vibrational levels $(v=0,L)$ and $(v=1,L)$ with $L=0,2$ (from Table~\ref{tab-probapair}). The spectrum is centered around the spin-independent transition frequency given in Table~\ref{spinless}.}
\end{figure}
\begin{figure}
\includegraphics[width=12cm,angle=-90]{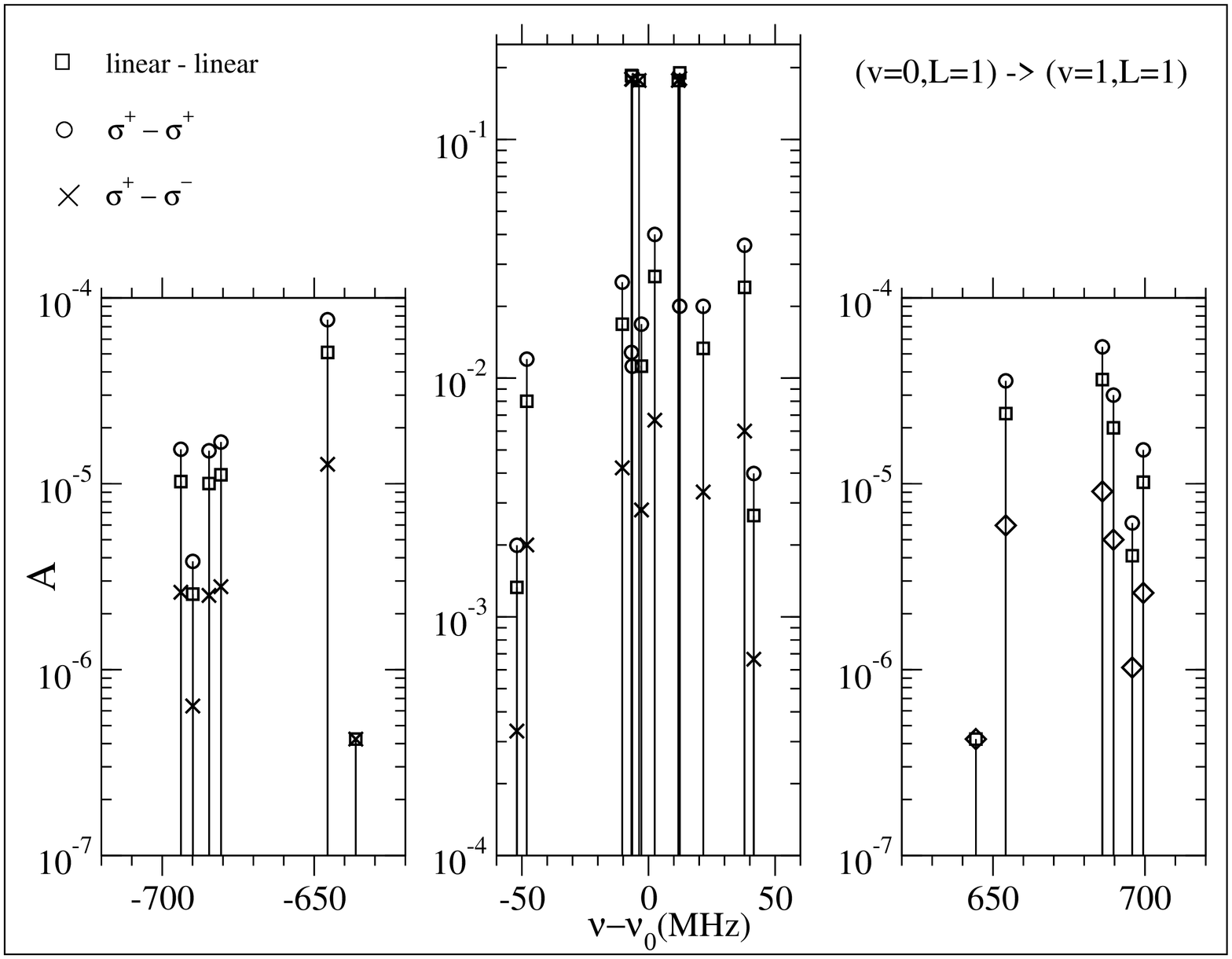}
\caption{\label{fig-probaimpair11} Same as Fig.~\ref{fig-probapair}, with $L=1$.}
\end{figure}
\begin{figure}
\includegraphics[width=12cm,angle=-90]{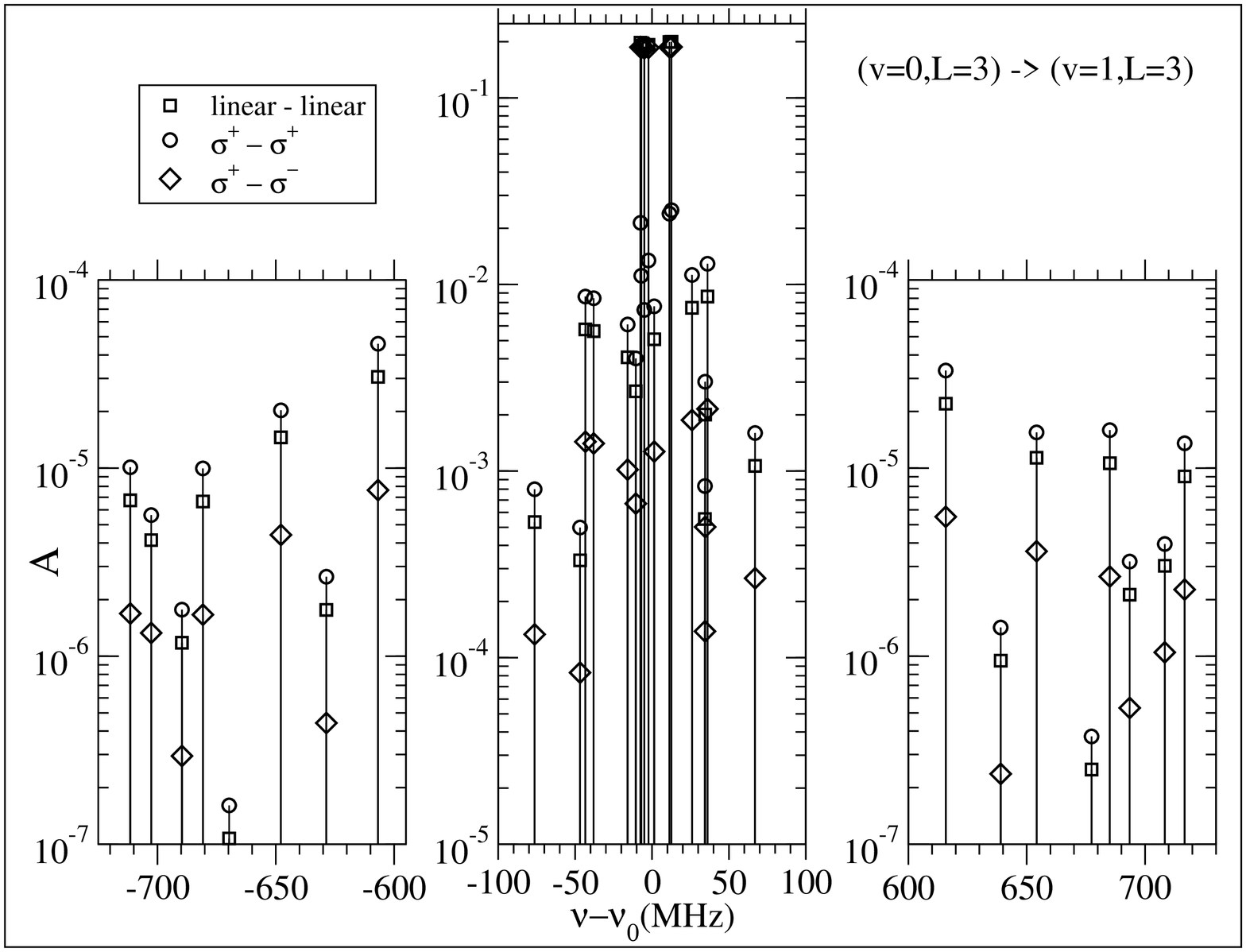}
\caption{\label{fig-probaimpair33} Same as Fig.~\ref{fig-probapair}, with $L=3$.}
\end{figure}
%%%%%%%%%%%%%%%%%%%%%%%%%%%%%%%%%%%%%%%%%%%%%%%%%%%%%%%%%%%%%%%%%%%%%%%%%%%%
Using Eqs.~(\ref{eqSQ2cb}), (\ref{eq6jgeneral}), the orbital reduced matrix elements given in Table~\ref{Q0Q2}, and the mixing coefficients given in Table~\ref{tab-shiftimpair}, we have computed the two-photon matrix elements for the four transitions $(v\!=\!0,L) \rightarrow (v'\!=\!1,L)$ with $0 \leq L \leq 3$. They are given in Tables~\ref{tab-probapair}, \ref{tab-probaL1} and~\ref{tab-probaL3} for $L=0$ and $2$, $L=1$ and $L=3$ respectively. The corresponding spectra, for three different choices of standard polarizations: linear-linear, $\sigma^+ \sigma^+$ and $\sigma^+ \sigma^-$, are shown in Figs.~\ref{fig-probapair}, \ref{fig-probaimpair11} and~\ref{fig-probaimpair33}.

Simplest is, of course, the $L\!=\!0$ case, where is no hyperfine splitting. The transition probability had been computed in Ref.~\cite{hilico2001} for linear-linear polarizations. Note that the transition is forbidden for $\sigma_+ \sigma_+$ polarizations, because of the selection rule $\Delta M_J\!=\!2$. In the $L\!=\!2$ case, there are two intense $\Delta J\!=\!0$ lines shifted by a few MHz and two weak $\Delta J\!=\!\pm 1$ lines shifted by about 50 MHz.

The spectra are more complex for odd values of $L$. They consist in one main cluster of intense $\Delta F\!=\!0$ lines which is about 50-100~MHz wide, and two satellite clusters of very weak lines (corresponding to $\Delta F\!=\!\pm 1$) about 600-700~MHz away. The total number of lines is 25(34) for $L=1$($3$) but the most intense are those of $\Delta F\!=\!\Delta J\!=\!0$; there are 5(6) of them for $L=1$($3$).

Whatever the value of $L$, all the favored transitions are between states with similar spin structure (i.e. same values of $F,J$). This feature makes them especially attractive for metrological purposes. Indeed, in such pairs of homologous hyperfine states, systematic shifts like the Zeeman shift (see Ref.~\cite{paper2}) are expected to have similar values, so that the shift of the transition frequency will be much smaller. The same is true for hyperfine structure corrections to the transition frequency: as can be seen e.g. from Figs.~\ref{fig-probaimpair11} and~\ref{fig-probaimpair33}, the most intense lines span a frequency interval of less than 25 MHz because the spin-dependent corrections to the initial and final state energies partially cancel each other. For this reason, the theoretical uncertainty on the frequency of these transitions is much smaller with respect to the other ones. On the whole, the favored transitions benefit at the same time from a smaller sensitivity to systematic effects, and from potentially more accurate theoretical predictions.

%%%%%%%%%%%%%%%%%%%%%%%%%%
\subsection{Orders of magnitude}

\begin{figure}
\includegraphics[width=14cm]{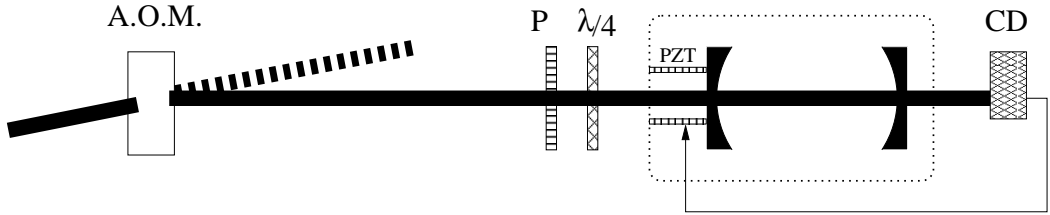}
\caption{\label{fig-sigma} Setup of our experiment for excitation with circular polarizations. A.O.M. stands for acousto-optic modulator, $P$ for polarizer, CD for cold detector.}
\end{figure}

The two-photon transition probability at resonance is
\begin{equation}
\Gamma=\left(\frac{4\pi a_0^3}{\hbar c}\right)^2 \frac{4}{\Gamma_f} I^{2} \left[ ^S\bar{Q}_{q_1,q_2} \right]^2 \label{probares}
\end{equation}
where $a_0$ is the Bohr radius, $\Gamma_f$ the instrumental width of the transition, and $I$ is the laser beam intensity. The above results show that the averaged two-photon matrix element for the favored transitions does not depend critically on the value of $L,F,J$ but strongly depends on the excitation polarizations; we have the typical values $\left[ ^S\bar{Q}_{q_1,q_2} \right]^2 \sim$ 0.2 for the case of linear-linear or $\sigma_+ \sigma_-$ polarizations and $\left[ ^S\bar{Q}_{q_1,q_2} \right]^2 \sim$ 0.02 for the $\sigma_+ \sigma_+$ case.
In this subsection we evaluate the two-photon transition probability, using the parameters of our experiment. Our excitation source is a QCL phase-locked to a CO$_2$ laser~\cite{bielsa2007}, which delivers a linearly polarized beam with a cw power of about 90~mW. A Fabry Perot cavity of finesse 1000 is built around the ion cloud. The QCL requires a strong optical isolation due to its extreme sensitivity to optical feedback from the high finesse Fabry Perot cavity. An optical isolation of more than 23dB (with 90\% transmission) can be achieved using an optical diode made of a grid polariser and a quarter-wave plate, which implies working with $\sigma_+ \sigma_+$ polarizations; the isolation ratio is limited by the polarizer extinction ratio~\cite{mansfield1980}. An additional isolation of 6 dB is obtained using an acousto-optic modulator with a polarization-dependent efficiency. The setup we have implemented is shown in Figure~\ref{fig-sigma}. The overall transmission of those optical elements including the alignment mirrors is 60\% so that 54~mW of optical power are injected into the high finesse cavity. The transmitted power at resonance is about 10~mW; from transmission and reflectivity measurements we estimate the mirror transmission and losses to about 0.001, so that the incident power on the H$_2^+$ ions is then $P \sim$~10W in a beam of waist $w_0 =$~1mm. The intensity on the beam axis is $2 P/\pi w_0^2 \sim 6.4$ W.mm$^{-2}$. The instrumental width is essentially the laser width $\Gamma_f =$ 2$\pi \times$ 2.6~kHz~\cite{ThBielsa}, since the width of the excited state is extremely small, all the ro-vibrational states of H$_2^+$ being metastable. From equation~(\ref{probares}) one obtains a transition probability $\Gamma \sim$ 0.7 s$^{-1}$ with $\sigma_+ \sigma_+$ polarizations.

The transition probabilities are higher by about one order of magnitude in the linear-linear polarization case. Due to this, even with a circularly polarized beam it is still more advantageous to probe the $\Delta M_J = 0$ transitions in a transverse magnetic field, which must be sufficiently strong to separate the three components $\Delta M_J = 0, \pm 2$. A field in the 100 mG -- 1 G range is enough, as estimated in~\cite{paper2}. The incident intensity is then decomposed into 50\%, 25\% and 25\% of linear, $\sigma_-$ and $\sigma_+$ polarizations respectively. A factor of 2 is lost on the excitation beam intensity (hence 4 on the transition probability), but this is more than compensated by the difference in the two-photon matrix element. With these parameters, the transition probability is $\Gamma \sim$ 1.7 s$^{-1}$, a large enough value to observe a two-photon transition in Paul traps where the ion lifetime is typically of several seconds. Further improvement can be achieved either by a tighter focusing of the laser (and a smaller ion cloud section in order to minimize transit-time broadening) or by reducing the laser linewidth.

\begin{figure}
\includegraphics[width=14cm]{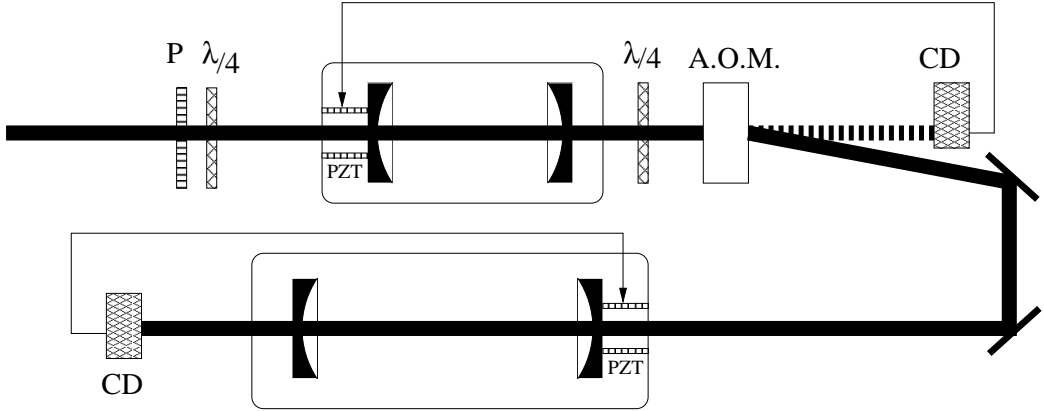}
\caption{\label{fig-linear} Proposed experimental setup for excitation with linear polarizations. A.O.M. stands for acousto-optic modulator, $P$ for polarizer, CD for cold detector.}
\end{figure}

A factor of 4 on the transition probability can be gained by using a linearly polarized excitation beam. In this case, the standard optical isolation technique relies on Faraday isolators. In the 9 micron range, a 45° polarization rotation with reasonable magnetic fields can only be obtained using n-doped InSb wafers under cryogenic conditions, with high insertion losses~\cite{dennis1967,jacobs1974}. In addition, to our knowledge they are no longer commercially available. Other ways of achieving isolation must be sought. Figure~\ref{fig-linear} shows our proposal for a high-transmission high-isolation device for linear polarization. It takes advantage of both constructive and destructive interference by a Fabry Perot cavity.

A Fabry Perot cavity of free spectral range 4$f$ is locked on resonance with the laser of frequency $\nu_L$, resulting in a high transmission. The transmitted beam is frequency shifted to $\nu_L+f$ by an acousto-optic modulator driven at a frequency $f$, and injected into the high finesse Fabry Perot cavity surrounding the ion cloud. On the way back to the QCL, the reflected beam is diffracted again by the acousto-optic modulator and shifted to $\nu_L+2f$. It is then exactly off resonance with the first Fabry Perot cavity that provides optical isolation. To summarize, the high transmission is due to constructive interference and isolation to destructive interference. Optical isolation of the QCL against the feedback from the first cavity can be achieved using an optical diode as discussed above. A second quarter-wave plate turns the polarization back to linear at the output of the isolation cavity.

The performances of this setup can be estimated as follows. The transmission at resonance for a Fabry Perot cavity made of two identical mirrors of reflectivity $R$, transmission $T$ and losses $P$ with $R+T+P=1$, is
\begin{equation}
T_{cav}=\frac{1}{\left[1+P/(1\!-\!R\!-\!P)\right]^2},
\end{equation}
and the off resonance isolation ratio expressed in dB is given by
\begin{equation}
I = -10\; \log_{10}\left[ \frac{(1-R-P)^2}{(1+R)^2}\right].
\end{equation}
Using low-losses mirrors with $R=$0.98 and $P=$0.001, one obtains $T_{cav}=$0.9 and $I=$40 dB.

\section{Conclusion}

We have presented a derivation of the hyperfine structure of two-photon transition spectra in the H$_2^+$ molecular ion, and applied it to several rotational components of the fundamental vibrational transition $(v\!=\!0,L) \rightarrow (v'\!=\!1,L)$. It was shown that the most intense lines are those between pairs of homologous hyperfine states $(v,L,F,J) \rightarrow (v',L,F,J)$. Our estimate reveals that observation of such lines in Doppler-free spectroscopy is feasible with present-day laser sources. We have also proposed an experimental setup allowing to probe the two-photon transitions with linear-linear polarizations. Let us point out that the experimental task of finding the transition frequency is made easier by recent progress in theoretical predictions~\cite{korobov2006b}. The current theoretical uncertainty on the spin-independent frequencies (given in Table~\ref{spinless}) is about 13 kHz~\cite{korobov}, while the uncertainty due to hyperfine corrections is of the order of 5 kHz, due to partial cancellation between the shifts of initial and final states. We have also shown that such transitions have a very low sensitivity to external magnetic fields~\cite{paper2}.
\begin{table}
\begin{tabular}{|@{\hspace{3mm}}c@{\hspace{3mm}}|@{\hspace{3mm}}c@{\hspace{3mm}}|@{\hspace{3mm}}c@{\hspace{3mm}}|}
\hline
$L$& $\nu_{2ph}$ (MHz) & $\lambda_{2ph}$ ($\mu$m)\\
\hline
0& 32 844 161.844 & 9.128 \\
1& 32 798 213.622 & 9.141 \\
2& 32 706 607.796 & 9.166 \\
3& 32 569 919.581 & 9.205 \\
\hline
\end{tabular}
\caption{\label{spinless}Spin-independent frequency and wavelength of the $(v\!=\!0,L) \rightarrow (v\!=\!1,L)$ transitions, with $0 \leq L \leq 3$. They were calculated using the data of Refs.~\cite{korobov2006b,korobov}.}
\end{table}

This work was supported by l'Universit\'e D'Evry Val d'Essonne and by la R\'egion Ile-de-France. V.I.K.\ acknowledges support of the Russian Foundation for Basic Research under Grant No.\ 08-02-00341. Laboratoire Kastler Brossel de l'Universit\'e Pierre et Marie Curie et de l'Ecole Normale Sup\'erieure is UMR 8552 du CNRS.
%%%%%%%%%%%%%%%%%%%%%%%%%%%%%%%%%%%%%%%%%%%%%%%%%%%%%%%%%%%%%%%%%%%%%%%%%%

\end{document}